\documentclass[runningheads]{llncs}

\usepackage{graphicx}
\usepackage{amssymb}
\usepackage{amsmath}
\usepackage{hyperref}
\usepackage{multirow}
\usepackage{makecell}
\usepackage{enumerate}
\usepackage[inline]{enumitem}
\usepackage[caption=false]{subfig}
\usepackage{booktabs}

\input{macros-generic}
\newcommand{\Prot}{\Pi}
\newcommand{\Exec}{\mathcal{E}}
\newcommand{\Honest}{\mathcal{H}}
\newcommand{\slot}{sl}
\newcommand{\epoch}{ep}
\newcommand{\party}{\mathsf{P}}
\newcommand{\Chain}{\mathsf{C}}
\newcommand{\Parties}{\mathcal{P}}
\newcommand{\adv}{\mathcal{A}}
\newcommand{\SD}{\dist{SD}}
\renewcommand{\SS}{\mathsf{SS}}
\newcommand{\SSl}{\hat{\mathsf{SS}}}
\newcommand\dist[1]{\mathcal{#1}}
\newcommand\set[1]{\mathcal{#1}}

\newcommand{\Lag}{\Lambda}
\newcommand{\Ledger}{\mathbf{L}}

\begin{document}
\title{Stake Shift in Major Cryptocurrencies: An Empirical Study}
\titlerunning{Stake Shift in Major Cryptocurrencies}
%
\author{
    Rainer St\"u{}tz\inst{1}\orcidID{0000-0001-9244-1441} \and
    Peter Ga\v{z}i\inst{2} \and
    Bernhard Haslhofer\inst{1}\orcidID{0000-0002-0415-4491} \and
    Jacob Illum\inst{3}
}
\authorrunning{R. St\"u{}tz et al.}
\institute{
    AIT Austrian Institute of Technology \and
    IOHK \and
    Chainalysis
}


%
\maketitle              
%

\begin{abstract}
In the proof-of-stake (PoS) paradigm for maintaining decentralized,
permissionless cryptocurrencies, Sybil attacks are prevented by basing the
distribution of roles in the protocol execution on the stake distribution
recorded in the ledger itself. However, for various reasons this distribution
cannot be completely up-to-date, introducing a gap between the present stake
distribution, which determines the parties' current incentives, and the one
used by the protocol.

In this paper, we investigate this issue, and empirically quantify its effects.
We survey existing provably secure PoS proposals to observe that the above
time gap between the two stake distributions, which we call \emph{stake
distribution lag}, amounts to several days for each of these protocols. Based
on this, we investigate the ledgers of four major cryptocurrencies (Bitcoin,
Bitcoin Cash, Litecoin and Zcash) and compute the average \emph{stake shift}
(the statistical distance of the two distributions) for each value of stake
distribution lag between 1 and 14~days, as well as related statistics. We also
empirically quantify the sublinear growth of stake shift with the length of
the considered lag interval. 

Finally, we turn our attention to unusual stake-shift spikes in these
currencies: we observe that hard forks trigger major stake shifts and that
single real-world actors, mostly exchanges, account for major stake shifts in
established cryptocurrency ecosystems.
\end{abstract}



\section{Introduction}\label{sec:intro}


The introduction of Bitcoin~\cite{bitcoin} represented the first practically
viable design of a cryptocurrency capable of operating in the so-called
permissionless setting, without succumbing to the inherently threatening Sybil
attacks. In the decade following Bitcoin's appearance, cryptocurrencies
have arguably become the main use case of the underlying blockchain technology.
Most deployed cryptocurrencies such as Bitcoin are relying on \emph{proofs of work
(PoW)} to prevent Sybil attacks and provide a robust transaction
ledger. However, the PoW approach, also has its downsides, most
importantly the associated energy waste (see e.g.~\cite{digiconomist}).

A promising alternative approach to maintaining a ledger in a permissionless
environment is based on so-called \emph{proof of stake (PoS)}, where Sybil
attacks are prevented by, roughly speaking, attributing to each participant in
the consensus protocol a weight that is proportional to his stake as recorded in
the ledger itself. Several PoS protocols embracing this idea have been shown to
achieve provable security guarantees in various
models~\cite{algorand,C:KRDO17,EPRINT:BenPasShi16b,EC:DGKR18,CCS:BGKRZ18,EPRINT:BGKRZ19}.


More concretely, in all these PoS schemes, whenever a protocol
participant needs to be selected for a certain role in the protocol, he is
chosen  with a probability that is proportional to his stake share in some
\emph{stake distribution} $\SD$, by which we mean a record of ownership of all
the assets maintained on the ledger at a given time, allowing to determine what
proportion of this stake is in control by any given party. In other words, the
stake distribution is a snapshot of the ownership of the ledger-based
asset at a given time (for simplicity of exposition, we assume only
a single-asset ledger in this discussion).

Ideally, the selection of a party for any security-relevant role in the protocol
at time $t$
should be based on a stake distribution $\SD$ that is as up-to-date as possible.
However, for various security-related reasons that we detail in
Section~\ref{sec:pos},
the protocols cannot use the ``current'' distribution of assets $\SD_{t}$ and
are forced to use $\SD_{t-\Lag}$ 
that
is recorded in the ledger up
to the point in time $t-\Lag$ for some time interval $\Lag$ that we
call the \emph{stake distribution lag} of the protocol.
However, roughly speaking, the security of the protocol is
determined by---and relies on a honest-majority assumption about---the
present stake distribution $\SD_{t}$. To account for this
difference, 
existing protocols assume that not too much money has changed hands
during the past time interval of length $\Lag$, and hence 
the distributions $\SD_{t-\Lag}$ and 
$\SD_{t}$ are close. 
Their distance, called \emph{stake shift}
in~\cite{C:KRDO17},
is the focus of our  present investigation.


\heading{Our Contributions.}

Up until now, the notion of stake shift has only been discussed on a theoretical
level and not yet quantified based on real-world data; 
we set up to fill this gap.
We conjecture that the
stake shift statistics of a cryptocurrency are mostly influenced by its
proliferation, market cap and daily trading volumes, rather than its underlying
consensus algorithm. Therefore, in an effort to understand the stake shift
characteristics of a mature cryptocurrency, we focus our analysis 
on PoW ledgers with a strong market dominance such as Bitcoin.\footnote{%
As of September 13, 2019, about $68\%$ of the total market capitalization of
cryptocurrencies is stored in Bitcoin (cf.\ \url{https://coinmarketcap.com}).}
We perform a systematic, empirical study of the stake shift phenomenon. 
More concretely, our contributions can be summarized as follows: 

\begin{enumerate}

\item 
We adjust the formal definition of stake shift given in~\cite{C:KRDO17}
to be applicable to studying the execution of the protocol in
retrospect, based only on the stabilized ledger produced, without access to the
states held by the parties during its execution.

\item
We provide a scalable algorithmic method for computing stabilized stake shift
over the entire history of PoW ledgers following the UTXO model. We computed
it in ledgers of four major cryptocurrencies (Bitcoin, Bitcoin Cash, Litecoin,
and Zcash) from their inception until July 31\textsuperscript{st}, 2019.

\item 
We study the evolution of stabilized stake shift in all ledgers and found
that hard forks may trigger major stake shifts. We also fitted a simple
quadratic polynomial model that mimics the real-world sublinear growth of
stake shift with respect to the considered stake distribution lag.

\item
We pick top spikes occurring within the last two years, and determine the
likely real-world identities behind them. We can observe that exchanges
are behind those spikes, at least in established cryptocurrencies such as
Bitcoin or Bitcoin Cash.

\end{enumerate}


Our results show that the stake-shift phenomenon has a noticeable impact on the
provable-security guarantees provided by PoS protocols from the literature. 
We argue in Section~\ref{sec:pos} that the stake shift over the stake distribution
lag period of a PoS protocol counts directly against the threshold of
adversarial
stake it can tolerate (typically 1/2 or 1/3), and the values of stake shift that
we observe are clearly significant on this scale, as we detail in
Section~\ref{sec:discussion}.

While our initial intention was to inform the design of PoS protocols, we believe
that our results can be interesting to a wider community and shed some light on
the real-life use of the studied cryptocurrencies as tools for value transfer.
Therefore, we make our research reproducible by releasing the implementation of our
stake shift computation method. It can be
used for computing stabilized stake shifts with configurable lag for any other
cryptocurrency that follows Bitcoin's UTXO model.

Finally, note that all measurements were performed on UTXO-based currencies
and some of the mentioned PoS protocols envision an account-based ledger. This
aspect, however, is completely irrelevant to our investigation.
Also, while our motivation comes from PoS protocols, we believe 
that most robust and useful data can be obtained from
mature blockchains and hence we focus our measurements on PoW ledgers. To
reemphasize, it seems reasonable to believe that the maturity (age, market cap,
trading volume, etc.) of a blockchain are more determining for its stake shift
behavior than the underlying consensus mechanism, hence justifying our choice. 

We start by providing more details on the relevance of stake shift for PoS
security, and
survey the stake distribution lags in existing proof-of-stake protocols
in Section~\ref{sec:background}. Then we provide a formal definition of stabilized
stake shift in Section~\ref{sec:stake_shift} and describe our datasets
and computation methods in Section~\ref{sec:methodology}. We present our findings
in Section~\ref{sec:results} and discuss them in Section~\ref{sec:discussion}.


\section{Background}\label{sec:background}

In this section we provide a more detailed discussion of the relevance of
stake shift for PoS protocols, and survey stake distribution lags of
several known PoS proposals.

\subsection{Importance of Stake Shift for Security of PoS-Based Blockchains}
\label{sec:pos}

As mentioned in Section~\ref{sec:intro},
the selection of a party for any security-relevant role in a PoS protocol
should ideally be based on a stake distribution $\SD$ that is as up-to-date as possible.
However, this is often difficult, as we detail next.

First, in the eventual-consensus PoS protocols such
as~\cite{C:KRDO17,EPRINT:BenPasShi16b,EC:DGKR18,CCS:BGKRZ18,EPRINT:BGKRZ19}, there is no
consensus about the inclusion of the most recently created blocks into the stable
ledger, such a consensus is only achieved gradually by adding more and more
blocks on top of them.
Consequently, during the protocol execution, the view of the current stake distribution
$\SD_t$ at time $t$ by different honest parties might actually differ and hence
$\SD_t$ cannot be used for electing protocol actors.

On the other hand, in PoS protocols based on Byzantine Agreement
such
as~\cite{algorand},
where the consensus about a block is achieved before proceeding to further blocks,
the most recent stake distribution still cannot be used for sampling protocol
participants. The reason is that the security of the protocol requires the stake
distribution to be old enough so that it was fully determined \emph{before} the
adversary could have any information about the bits of randomness used to sample
from this distribution (which are also produced by the protocol).

Therefore, in all these protocols, participants that are allowed to act at some
time $t$ are sampled according to a distribution $\SD_{t-\Lag}$
recorded in the ledger up to the point in time $t-\Lag$ for some
stake distribution lag $\Lag$.
This is done with the intention that $\SD_{t-\Lag}$ is both
\begin{itemize}
  \item stable (in the case of eventual-consensus protocols), and
  \item recent enough so that it can be assumed that it does not
differ too much from the current distribution $\SD_t$.
\end{itemize}

However, the incentives of the participants are, of course, shaped by the current
distribution of the stake: For example, a party $\party$ that used to own a
significant portion of the stake, but has just transferred (e.g., sold) all of
it in time $t_1$, has no longer any stake in the system and hence no direct motivation to
contribute to its maintenance. Nonetheless, at any time $t$ during the time interval
$(t_1,t_1+\Lag)$, the stake distribution $\SD_{t-\Lag}$ will still attribute some stake to
$\party$ and hence $\party$ will be allowed (and expected) to act accordingly in the protocol.
This discrepancy is present in all PoS protocols listed above, and in fact in
all provably-secure PoS protocols in the literature.

The security of these PoS protocols is typically argued based on the assumption
that at any point during the execution, less than a fraction $T$ of the total
stake in the system is controlled by adversarial parties (for
$T=1/2$
in~\cite{C:KRDO17,EPRINT:BenPasShi16b,EC:DGKR18,CCS:BGKRZ18,EPRINT:BGKRZ19}
and $T=1/3$ in~\cite{algorand}).
To formally account for the above mismatch, one has to choose between the following
two approaches:
\begin{enumerate}[label=(\roman*)]
\item
  \label{assum1}
Directly assume that, at every point $t$ during the execution,
    less than a $T$-fraction of stake in the \emph{old} distribution $\SD_{t-\Lag}$ is controlled by
    parties that are adversarial \emph{at time $t$}.
\item
  \label{assum2}
Make an additional assumption that, at any point $t$ during the execution,
some (normalized) ``difference'' between $\SD_{t-\Lag}$ and the
current factual distribution of stake $\SD_{t}$  in the system is bounded by a
constant $\sigma\in(0,1)$; i.e., that not too much money has changed hands
    between $t-\Lag$ and $t$.
This assumption allows to conclude
    security as long as the
    \emph{current} adversarial stake ratio $\alpha\in[0,1]$ in $\SD_t$ satisfies
\begin{equation}
\label{eq:alpha}
\alpha
\leq
  (1-\eps)
  \cdot T
- \sigma
\end{equation}
    for some $\eps\geq 0$ (see e.g.~\cite[Theorem~6]{C:KRDO17}, respectively
    Theorem~5.3 in the full version of~\cite{C:KRDO17}).
\end{enumerate}

All of the provably secure PoS protocols adopt one of these two approaches.
While the assumption in approach~\ref{assum1} is
formally sufficient, it is arguably cumbersome and counter-intuitive, making
the reasoning~\ref{assum2} preferable.
As evidenced by equation~(\ref{eq:alpha}), in the approach~\ref{assum2} the
quantity $\sigma$, called stake shift,
plays a significant role for the protocols' security.

Let us clarify that our primary motivation for investigating stake shift
pertains to the distributions $\SD_t$ and $SD_{t-\Lag}$ as described above and
defined by individual PoS protocols, and does not aim at addressing the dangers
of long-range attacks (see e.g.~\cite{EPRINT:GazKiaRus18} for an overview of
those).
In a typical long-range attack setting, the
considered time interval would be much longer and one could hardly
expect a limited stake shift over it.

Finally, following the above motivation, below we focus on provably secure
PoS proposals. All these protocols use all existing coins for staking, not
distinguishing between ``staked'' and ``unstaked'' coins, and so we don't
consider this distinction below. It is worth mentioning
that practical implementations of these protocols, as well as other PoS
blockchains such as
EOS\footnote{\url{https://eos.io}}
and Tezos\footnote{\url{https://tezos.com}}, often deviate from this
approach and allow for coins that do not participate in staking.

\subsection{Stake Distribution Lag in Existing PoS Protocols}
\label{sec:stake_shift_proto}

Here we survey the value of stake distribution lag in several provably secure PoS protocol
proposals.

\heading{Ouroboros.}
The Ouroboros PoS protocol~\cite{C:KRDO17} divides its execution into so-called
\emph{epochs}, where each epoch is a sequence of $10k$ slots for a parameter
$k$ (this structure is dictated by the inner workings of the protocol). The stake
distribution used for sampling slot leaders in epoch $\epoch_j$ is the one reflected in
the current chain up to slot $4k$ of the preceding epoch $\epoch_{j-1}$.
Therefore, the stake distribution lag amounts to at most $14k$ slots.

In the deployment of the Ouroboros protocol in the Cardano
project~\footnote{\url{https://www.cardano.org}}, each
slot takes $20$ seconds and $k$ is chosen to be $2160$. Therefore, the above
upper bound on the stake distribution lag corresponds to exactly $7$~days.

\heading{Ouroboros Praos and Ouroboros Genesis.}
These protocols, which are defined in~\cite{EC:DGKR18,CCS:BGKRZ18}, also divide
their execution into epochs.  However, the stake distribution used for sampling
slot leaders in epoch $\epoch_j$ is the one reflected in the current chain up
to the last slot of the epoch $\epoch_{j-2}$. Hence the stake distribution lag
amounts to at most $2$~epochs.
Assuming the same epoch length as above, this would result in a stake
distribution lag of exactly $10$~days.

\heading{Algorand and Vault.}
For the protocols Algorand~\cite{algorand,EPRINT:GHMVZ17,EPRINT:CGMV18} and
Vault~\cite{EPRINT:LSGZ18}
we consider the parametrization given
in~\cite{EPRINT:LSGZ18}, where the authors suggest to consider a stake distribution
lag of $1$~day for Algorand and hence $2$~days for Vault.

\heading{Snow White.}
The Snow White protocol employs a ``look-back'' of $2\omega$~blocks for a
parameter $\omega$ that is sufficient to invoke the common-prefix and
chain-quality properties (see~\cite{EPRINT:BenPasShi16b}).
The authors do not propose a concrete value of $\omega$, however, given that the
requirements put on $\omega$ are similar to other protocols (common prefix,
chain quality), it is safe to assume that an implementation of Snow White would
also lead to a stake distribution lag between 1 and 10~days.

\section{Stabilized Stake Shift Definition}\label{sec:stake_shift}



We are interested in executions of blockchain ledger protocols, and will be assuming a
model in the spirit of~\cite{EC:GarKiaLeo15} to formalize such executions.
In particular, we assume there is an \emph{environment} orchestrating the
execution, a set of \emph{parties} $\Parties$ executing the protocol, and an
\emph{adversary}
$\adv$ allowed to corrupt the parties upon approval from the environment;
parties yet uncorrupted are called \emph{honest}.
We assume that the protocol execution is divided into a sequence of disjoint,
consecutive time intervals called \emph{slots}, indexed by natural numbers
(starting with $1$).
The set of honest parties at each slot $\slot$ is denoted by $\Honest[\slot]$.
Finally, we denote by $\Chain^\party[\slot]$ the chain held by an honest party
$\party$ at the beginning of slot $\slot$.

Finally, let $\SD^\party[\slot]$ denote the stake distribution
recorded in the chain $\Chain^\party[\slot]$
up to slot $\slot$, seen as a probability distribution (i.e., normalized to sum
to $1$).
As a notational convenience, let $\SD^\party[0]$ denote
the initial stake distribution recorded in the genesis block.

To define stake shift, we use the standard notion of statistical distance of two
discrete probability distributions.

\begin{definition}[Statistical distance]\label{def:stat-dist}
  For two discrete probability distributions $\dist{X}$ and $\dist{Y}$ with
  support $\set{S}_{\dist{X}}$ and $\set{S}_{\dist{Y}}$ respectively, the
  \emph{statistical distance} (sometimes also called the \emph{total variation
  distance})
  of $\dist{X}$ and $\dist{Y}$
  is defined as
$$
  \delta(\dist{X},\dist{Y})
  \defeq
  \tfrac{1}{2}\sum_{s\in\set{S}_{\dist{X}}\cup\set{S}_{\dist{Y}}}
    \left|
      \Pr_{\dist{X}}[s]
      -
      \Pr_{\dist{Y}}[s]
    \right|.
$$
\end{definition}

Seeing stake distributions as probability distributions allows for the
following definition inspired by~\cite[Definition 5.1]{C:KRDO17}.

\begin{definition}[Stake shift]
  Consider an execution $\Exec$ of a block\-chain protocol $\Prot$
  for $L$ slots, and let $\slot\in\{\Lag,\ldots,L\}$.
  The \emph{$\Lag$-stake shift} in slot $\slot$ is the maximum, over
  all parties $\party_1$ honest in slot $\slot-\Lag$ and
  all parties $\party_2$ honest in slot $\slot$,
  of the statistical distance between the stake
  distributions in slots $\slot-\Lag$ and $\slot$ as perceived by $\party_1$ and
  $\party_2$, respectively.
  Formally,
  $$
  \SS_\Lag(\Exec,\slot)
  \defeq
  \max_{
    \substack{
    \party_1\in\Honest[\slot-\Lag]\\
    \party_2\in\Honest[\slot]
  }
  }
  \delta\left(
    \SD^{\party_1}[\slot-\Lag],
    \SD^{\party_2}[\slot]
  \right).
  $$
  Naturally, we extend this notion over the whole execution and define
  the \emph{$\Lag$-stake shift} of $\Exec$ to be
  $$
  \SS_\Lag(\Exec)
  \defeq
  \max_{\Lag\leq\slot\leq L}
    \SS_\Lag(\Exec,\slot).
  $$
\end{definition}

Finally, note that the quantity $\SS_{\Lag}(\Exec,\slot)$, and consequently also
$\SS_{\Lag}(\Exec)$, cannot be determined based solely on the final stabilized
ledger $\Ledger$ that was created by the protocol, as it does not contain the views of the
participants during the protocol execution. For this reason, any long-term
empirical study that is only based on the  preserved stabilized ledger
$\Ledger$ (e.g. the Bitcoin blockchain)
has to aim for an analogous quantity capturing stake shift in $\Ledger$, as defined next.

For a stable ledger $\Ledger$, we denote by $\SD^{\Ledger}[\slot]$ the stake
distribution as recorded in $\Ledger$ up to slot $\slot$.

\begin{definition}[Stabilized stake shift]
\label{def:SSS}
Consider an execution $\Exec$ of a block\-chain protocol $\Prot$
for $L$ slots, let $\Ledger$ denote the resulting stable ledger produced by
$\Prot$ during $\Exec$, and let $\slot\in\{\Lag,\ldots,L\}$.
The \emph{stabilized $\Lag$-stake shift} in slot $\slot$ is defined as
$$
  \SSl_\Lag(\Exec,\slot)
  \defeq
  \delta\left(
    \SD^{\Ledger}[\slot-\Lag],
    \SD^{\Ledger}[\slot]
  \right),
$$
and similarly, the \emph{stabilized $\Lag$-stake shift} of $\Exec$ is
  $$
  \SSl_\Lag(\Exec)
  \defeq
  \max_{\Lag\leq\slot\leq L}
    \SSl_\Lag(\Exec,\slot).
  $$
\end{definition}

For the reasons noted above, we will focus on \emph{stabilized} stake shift in our
empirical analysis; whenever we use the term \emph{stake shift} below, we refer to
its stabilized variant as per Definition~\ref{def:SSS}.




\section{Data and Methods}\label{sec:methodology}

Before we can empirically investigate stake shifts in deployed
cryptocurrencies, we first need to translate the definition of
\emph{stake shift} into a scalable algorithmic procedure
that can compute stake shift with configurable lags over a currency's
entire history, which in the case of Bitcoin spans more than 440M
transactions and 0.5B addresses.
In this section, we describe how we prepare the required datasets
from the underlying blockchains and the technical details
of our stabilized stake shift computation method.

\subsection{Dataset Preparation and Structure}\label{subsec:methodology_dataset}

We consider datasets from four different cryptocurrency ledgers: first, we take
\emph{Bitcoin (BTC)}, which is still the cryptocurrency with the strongest
market dominance. Additionally, we take three alternatives derived from the
Bitcoin Core code base: \emph{Bitcoin Cash (BCH)}, which is a hard fork from
the Bitcoin blockchain to increase the block size limit, which took effect in
August 2017; \emph{Litecoin (LTC)}, which was an early altcoin, starting in
October 2011, and is very similar to Bitcoin. The key differences to Bitcoin
are its choice of the proof-of-work algorithm (\emph{scrypt}) and the network's
average block creation time, which is roughly 2.5 minutes. Finally, we also
consider \emph{Zcash (ZEC)}, which is a cryptocurrency with enhanced privacy
features, initially released in October 2016. Zcash coins are either in a
transparent or a shielded pool.
The transparent (unshielded) pool contains ZEC in transparent addresses
(so-called \emph{t-addresses}). Due to the anonymity features in
Zcash, our analysis is limited to the transparent transactions in the
unshielded pool. However, as observed in~\cite{kappos}, a large proportion of
the activity on Zcash does not use the shielded pools. A summary of the used
datasets is provided in Table~\ref{tab:data-summary}.

\begin{table*}[h]
    \centering%
    \caption{Summary of considered cryptocurrency datasets.}%
    \resizebox{\textwidth}{!}{\begin{tabular}{@{\extracolsep{\fill}}lrrrrrr}
  \toprule
Currency & \# Blocks & Last timestamp & \# Txs & \# Addresses & \# Clusters & \# Entities \\ 
  \midrule
BTC & 588,007 & 2019-07-31 23:55:05Z & 440,487,974 & 540,942,127 & 50,162,316 & 260,182,367 \\ 
  BCH & 593,795 & 2019-07-31 23:54:09Z & 275,765,798 & 302,098,643 & 31,173,961 & 142,884,996 \\ 
  LTC & 1,677,479 & 2019-07-31 23:57:21Z & 36,009,400 & 44,256,812 & 3,052,978 & 23,304,076 \\ 
  ZEC & 577,390 & 2019-07-31 23:59:54Z & 5,052,970 & 3,488,294 & 206,506 & 1,680,481 \\ 
   \bottomrule
\end{tabular}
}%
    \label{tab:data-summary}%
\end{table*}

For each cryptocurrency ledger, we partition these addresses into
maximal subsets (clusters) that are likely to be controlled by the
same entity using the well-known and efficient mul\-ti\-ple-input
clustering heuristics~\cite{meiklejohn2013fistful}.
The underlying intuition is that if two addresses (e.g., $A_{1}$ and
$A_{2}$) are used as inputs in the same transaction while one of these
addresses along with another address (e.g., $A_{2}$ and $A_{3}$) are
used as inputs in another transaction, then the three addresses
($A_{1}$, $A_{2}$ and $A_{3}$) must somehow be controlled by the same
entity, who conducted both transactions and therefore possesses the
private keys corresponding to all three addresses. Being aware that this
heuristic fails when CoinJoin transactions~\cite{MoeserB2016JoinMeOnAMarket}
are involved, we filtered those transactions before applying the
multiple-input heuristics.

Before describing our stake shift computation method in more detail,
we introduce the following notation for key entities in our dataset:
we consider a blockchain
$\mathsf{B}_{t_{\text{end}}} = (\mathsf{A}, \mathsf{T})$ with its
associated set of addresses~$\mathsf{A}$ and set of transactions $\mathsf{T}$
at time $t_{\text{end}}$.

The multiple-input heuristics algorithm is applied to the complete transaction
dataset at time~$t_{\text{end}}$ to obtain a set of clusters
$\mathsf{C} = \{C_{1}, \ldots, C_{n_{c}}\}$.
Each cluster $C_{i}$ is represented by a set of addresses, where
$\lvert C_{i} \rvert \geq 2, \forall i \in \{1, \ldots, n_{c}\}$.
The set of entities $\mathsf{E}$ is represented by the union of $\mathsf{C}$
with the remaining single address clusters, i.e.,
$ \mathsf{E} = \mathsf{C} \cup 
  \{\{a\} \,\vert\,
  a \in \mathsf{A} \wedge \forall C \in \mathsf{C}: a \notin C \}
$.
The (cumulative) balance for entity $e \in \mathsf{E}$ at time $t$ is denoted
by $b_{e}^{t}$, and the total balance over all entities at time $t$ is given
by $b_{\text{total}}^{t} = \sum_{e \in \mathsf{E}} b_{e}^{t}$.

The last three columns in Table~\ref{tab:data-summary} show the number of
addresses in each ledger, the number of computed clusters, as well as the
number of entities holding the corresponding private keys of one or more
addresses.

For further inspecting the real-world identities behind entities causing major
stake shifts, we rely on Chainalysis\footnote{\url{https://www.chainalysis.com/}},
which is a proprietary online tool that facilitates the tracking of Bitcoin
transactions by annotating Bitcoin addresses with potential owners.

\subsection{Stake Shift Computation}

Given the dimensionality of our dataset, the challenge lies in finding a method
that follows Definition~\ref{def:SSS} and can compute the distances~$\delta$ in
a scalable, distributed and memory-efficient manner.

In a na\"ive approach one would calculate the cumulative balance for each
entity at every time period (e.g., days). The stake distribution is represented
by the relative frequencies, which are the result of dividing the cumulative
balances at time $t_{p}$ by the total balance~$b_{\text{total}}^{t_{p}}$. This
approach would result in huge temporary datasets that must be persisted in
memory for subsequent computation steps. For instance, for the computation of
the stabilized stake shift in Bitcoin, a grid of 3,862 $\times$ 260,182,367
(number of days $\times$ number of entities) data points needs to be cached,
which is computationally inefficient and hardly feasible in practice given
today's hardware limitations.

Therefore, we propose an iterator-based approach coupled with a custom
aggregation method, which can be executed on a distributed, horizontally
scalable data processing architecture: First, we join the transaction data with
the relevant entity information, and use the entity IDs for partitioning.
Then, for calculating the cumulative balances, we sort every partition by time
period. The iterator represents basically a loop over the grid of predefined
time periods for a given entity. Internally, we build up a data structure that
holds the following information in each iteration step:
\begin{enumerate*}[label=(\roman*)]
  \item entity ID $e$,
  \item time period $t_{p}$,
  \item the cumulative balance $b_{e}^{t_{p}}$,
  \item the contribution of the current entity to the stake distribution
    $ R_{e}^{t_{p}} = {b_{e}^{t_{p}}}/{b_{\text{total}}^{t_{p}}}$
    at time $t_{p}$; and
  \item the absolute difference of the stake distribution
        contributions at time $t_{p}$ and $t_{p-\ell}$:
        $\delta_{e}^{t_{p}} =
         \lvert R_{e}^{t_{p}} - R_{e}^{t_{p-\ell}} \rvert.$
\end{enumerate*}

To compute the stake shift for arbitrary lag values~$\ell$, a FIFO
(first in, first out) structure is needed to hold at most $\ell$ instances
of the above data structure for the last $\ell$ periods. 
That data structure can efficiently be partitioned
across computation nodes and requires zero communication costs.
An aggregation method then collects all partial results to obtain the
stake shift value~$\SSl_{\ell}^{t_{p}}$ at time period~$t_{p}$.

We implemented our stake shift computation method as single Apache
Spark\footnote{\url{https://spark.apache.org/}} job operating directly on 
a pre-computed dataset provisioned by the GraphSense Cryptocurrency Analytics
Platform\footnote{\url{https://graphsense.info}}. For further technical details,
we refer to the source code, which will be released with this paper.

\section{Analysis and Results}\label{sec:results}

In the following, we first report results on the longitudinal evolution of
stake shifts in all considered cryptocurrencies (BTC, BCH, LTC, ZEC).
Then we handpick past stake shift spikes and analyze them in more detail,
in order to gain a better understanding on the factors causing those shifts.
We also elaborate on cross-ledger similarities and differences.

\subsection{Evolution of Stabilized Stake Shifts}\label{sec:results_shifts}

\begin{figure}
  \includegraphics[width=\textwidth]{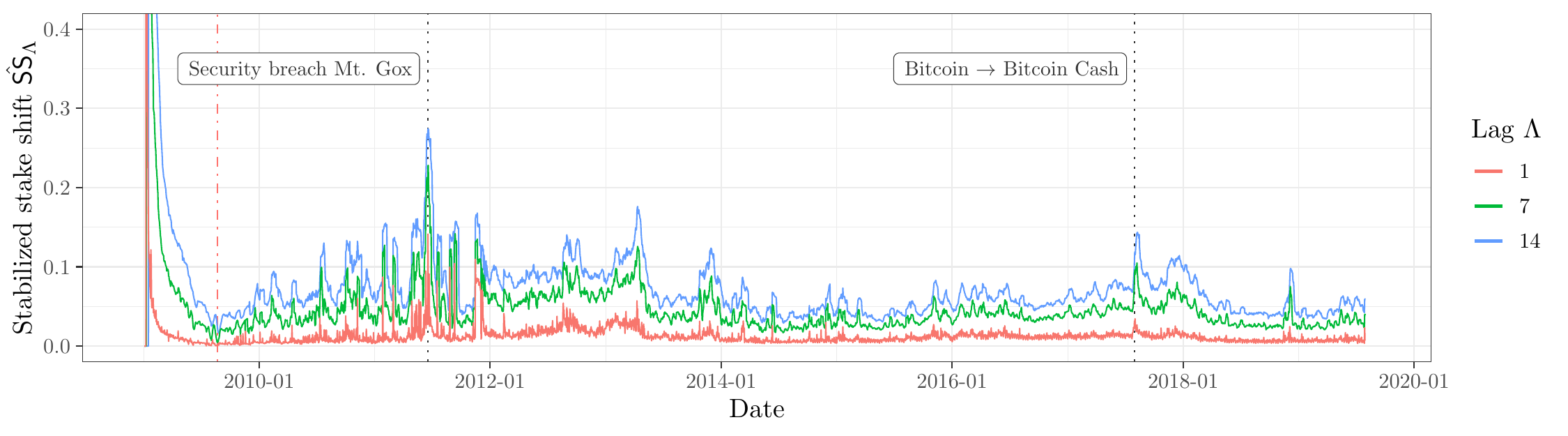}
  \caption{Stake shift for BTC (stake distribution lag $\Lag$: 1, 7, and 14 days).}
  \label{fig:stake_shift_lags_btc}
\end{figure}

Figure~\ref{fig:stake_shift_lags_btc} depicts the evolution of \textbf{Bitcoin}
stake shifts over the observation period for three different lag settings
$\Lag$: 1 day, 7 days, and 14 days. We can observe huge spikes
(0.933 for $\Lag = 1$) right after the generation of the genesis block and
another major spike occurring on June 19\textsuperscript{th}, 2011. That spike
is most likely related to a security breach on Mt.\ Gox, at this time one of
the dominating Bitcoin exchanges. After an attacker illegally transferred a
large amount of Bitcoins, 424,242~BTC were moved from a \emph{cold storage} to a
Mt.\ Gox address on June 23\textsuperscript{rd} 2011%
\footnote{\url{https://en.wikipedia.org/wiki/Mt._Gox}}.
We can also observe that hard forks trigger major stake shifts: Bitcoin Cash,
for instance, hard forked on August 1, 2017.

Due to the lack of space, we will in the following refrain from depicting 
stake evolutions for the other investigated currencies and focus on reporting
key observations and findings instead. For further visual inspection, we refer
the interested reader to the Appendix of this paper. We also restrict subsequent
discussions to $\Lag = 1$ because we can observe that stake shifts evolve
synchronously and differ only in lag amplitudes.

\textbf{Bitcoin Cash} shows similar behavior to Bitcoin: since it is a hard
fork of Bitcoin, stake shifts run synchronous to Bitcoin until the hard fork
date. Stake shift values in Bitcoin Cash also show a higher
variability after November 15, 2018. On this date a hard fork
was activated by \emph{Bitcoin~ABC}\footnote{\url{https://www.bitcoinabc.org/}}
(at the time the largest software client for Bitcoin Cash) and
\emph{Bitcoin~SV}\footnote{\url{https://bitcoinsv.io/}} (Satoshi's Vision).

In general, the variability of stake shifts in \textbf{Litecoin} (\$4.7B market
capitalization) appears to be higher than the one in Bitcoin. The biggest
spikes appear on the following dates: 2014-02-05, 2015-03-08, and 2018-11-30.
The first two spikes are represented by a couple of dominating entities.
We observed either a direct currency flow between them, or a indirect flow
via some intermediary cluster or address.
One exception is the spike on November 30\textsuperscript{th}, 2018: on that
day, approximately 35.4M~LTC were transferred within a 24 hour period, with a
total value of \$1.1B at that time. This is extraordinary, because the Litecoin
network has recorded approximately \$100M of trading volume per day, on
average.
After investigating involved transactions, we noted that a significant portion
of the transaction volume appears to originate from a single entity, which was
not captured by the multiple-input clustering heuristic. At least 40~new
wallets have entered the list of the richest Litecoin addresses, each with a
balance of 300,000~LTC ($\sim$\$10M).
In total, the addresses account for 12.9M~LTC (approximately \$372M). The
reason for the movement is still unclear, but, as we will discuss later in
Section~\ref{sec:results_spikes}, we can observe that the entities involved in
those stake shifts were controlled by Coinbase, which is a major cryptocurrency
exchange.

\begin{figure}
  \centering
  \includegraphics[width=\columnwidth]{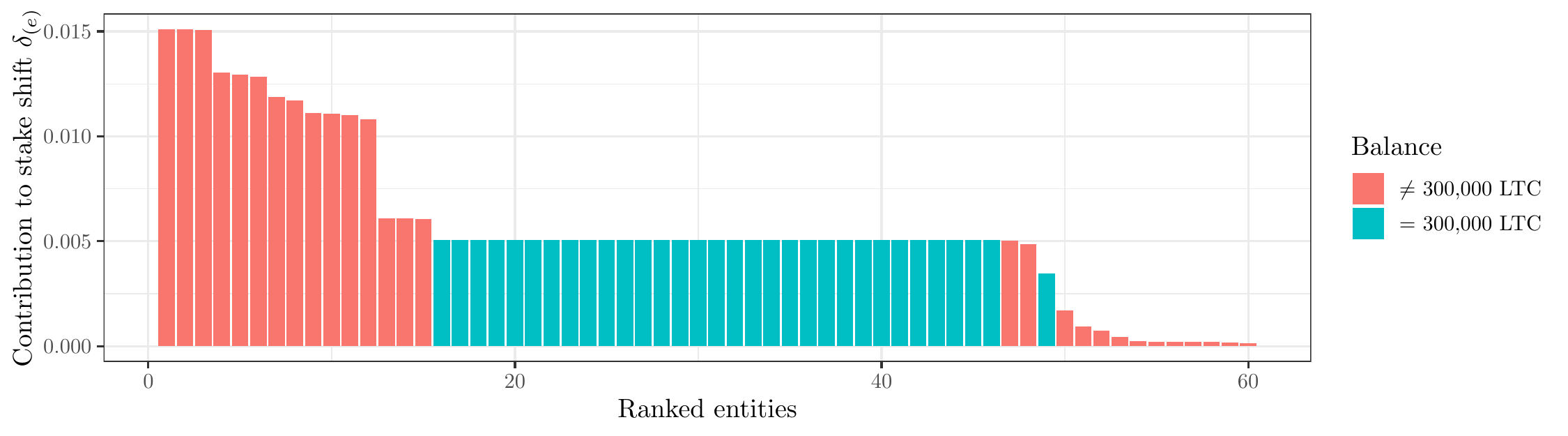}
    \caption{Ranked contributions (top 60) to stabilized stake shift for $\Lag=1$ 
             (LTC on November 30\textsuperscript{th}, 2018).}
  \label{fig:stakeshift_contribution_ltc17865}
\end{figure}

Figure~\ref{fig:stakeshift_contribution_ltc17865} provides a more detailed
view on that single Litecoin spike. It shows the top 60 contributions to the
stake shift for Litecoin on November 30\textsuperscript{th}, 2018. A block of
consecutive addresses sharing a certain transaction behavior becomes visible
between rank 16 to 46. They share the following common characteristics:
\begin{enumerate*}[label=(\roman*)]
  \item the number of incoming transaction is either 40 or 41;
  \item transactions are executed in chunks of 7,500 LTC; and
  \item the total balance is 300,000 LTC.
\end{enumerate*}

The remaining 11~addresses of this entity appear in the tail of the distribution.
The reason is that the transactions already started on the day before
(2018-11-29 21:18:59Z). Therefore, these 11~addresses do not (fully) account 
to the stabilized stake shift of November 30\textsuperscript{th}, 2018.

When regarding the stake shift evolution of \textbf{Zcash} (\$366M market
capitalization), we can, as in Litecoin, observe higher variability than in
Bitcoin or Bitcoin Cash. This could be explained by the differences in market
capitalization (\$5.5B BCH vs. \$177B BTC) in these two
currencies\footnote{\url{https://coinmarketcap.com/all/views/all/}, retrieved
on 2019-09-19.}.

\begin{table*}[h]
    \centering%
    \caption{Summary statistics of stabilized stake shift for different lag values.}%
    \resizebox{\textwidth}{!}{\begin{tabular}{@{\extracolsep{4pt}}rrrrrrrrrrrrr@{}}
  \toprule
&\multicolumn{3}{c}{BTC} & \multicolumn{3}{c}{BCH} & \multicolumn{3}{c}{LTC} & \multicolumn{3}{c}{ZEC}\\
 \cline{2-4} \cline{5-7} \cline{8-10} \cline{11-13}
\thead{Lag\\(in days)} & Mean & Median & Std Dev & Mean & Median & Std Dev & Mean & Median & Std Dev & Mean & Median & Std Dev \\ 
  \midrule
1 & 0.013 & 0.010 & 0.0098 & 0.013 & 0.011 & 0.0102 & 0.014 & 0.011 & 0.0123 & 0.014 & 0.012 & 0.0102 \\ 
  2 & 0.020 & 0.017 & 0.0129 & 0.020 & 0.017 & 0.0134 & 0.022 & 0.017 & 0.0177 & 0.023 & 0.020 & 0.0146 \\ 
  3 & 0.026 & 0.022 & 0.0155 & 0.026 & 0.023 & 0.0161 & 0.030 & 0.023 & 0.0219 & 0.031 & 0.027 & 0.0181 \\ 
  4 & 0.031 & 0.027 & 0.0177 & 0.032 & 0.027 & 0.0183 & 0.036 & 0.029 & 0.0255 & 0.038 & 0.034 & 0.0211 \\ 
  5 & 0.036 & 0.031 & 0.0196 & 0.037 & 0.032 & 0.0203 & 0.042 & 0.034 & 0.0289 & 0.045 & 0.040 & 0.0238 \\ 
  6 & 0.040 & 0.035 & 0.0213 & 0.041 & 0.036 & 0.0221 & 0.048 & 0.039 & 0.0319 & 0.051 & 0.047 & 0.0262 \\ 
  7 & 0.045 & 0.039 & 0.0229 & 0.045 & 0.039 & 0.0238 & 0.053 & 0.044 & 0.0347 & 0.058 & 0.053 & 0.0286 \\ 
  8 & 0.049 & 0.042 & 0.0244 & 0.050 & 0.043 & 0.0253 & 0.058 & 0.048 & 0.0374 & 0.063 & 0.059 & 0.0308 \\ 
  9 & 0.053 & 0.045 & 0.0257 & 0.053 & 0.046 & 0.0267 & 0.063 & 0.052 & 0.0399 & 0.069 & 0.065 & 0.0328 \\ 
  10 & 0.056 & 0.049 & 0.0270 & 0.057 & 0.050 & 0.0281 & 0.068 & 0.057 & 0.0423 & 0.074 & 0.070 & 0.0346 \\ 
  11 & 0.060 & 0.052 & 0.0282 & 0.061 & 0.053 & 0.0293 & 0.073 & 0.060 & 0.0446 & 0.079 & 0.075 & 0.0364 \\ 
  12 & 0.063 & 0.055 & 0.0294 & 0.064 & 0.056 & 0.0305 & 0.077 & 0.064 & 0.0469 & 0.084 & 0.081 & 0.0380 \\ 
  13 & 0.067 & 0.058 & 0.0305 & 0.068 & 0.059 & 0.0317 & 0.082 & 0.068 & 0.0490 & 0.089 & 0.085 & 0.0395 \\ 
  14 & 0.070 & 0.061 & 0.0316 & 0.071 & 0.062 & 0.0329 & 0.086 & 0.072 & 0.0510 & 0.094 & 0.090 & 0.0410 \\ 
   \bottomrule
\end{tabular}
}%
    \label{tab:stake_shift_stat}%
\end{table*}

More detailed statistics for stake distribution lag $\Lag$ ranging from 1 to
14~days are summarized in Table~\ref{tab:stake_shift_stat}, which shows the
mean, median, and standard deviation of resulting stake shift values.
Since the estimators for the arithmetic mean and standard deviation are not
robust against outliers, we did not consider the initial parts of the time
line and disregarded the first 6\% of the total number of days in our
estimation (marked with red dash-dotted vertical line in
Figure~\ref{fig:stake_shift_lags_btc} and Figure~\ref{fig:stake_shift_lags},
respectively). The gradually increasing mean and median stake shift values
confirm our previous observation of growing amplitudes.

\subsection{Modeling Stake Shift}\label{sec:shifts_model}

\begin{figure*}
  \centering
  \includegraphics[width=\textwidth]{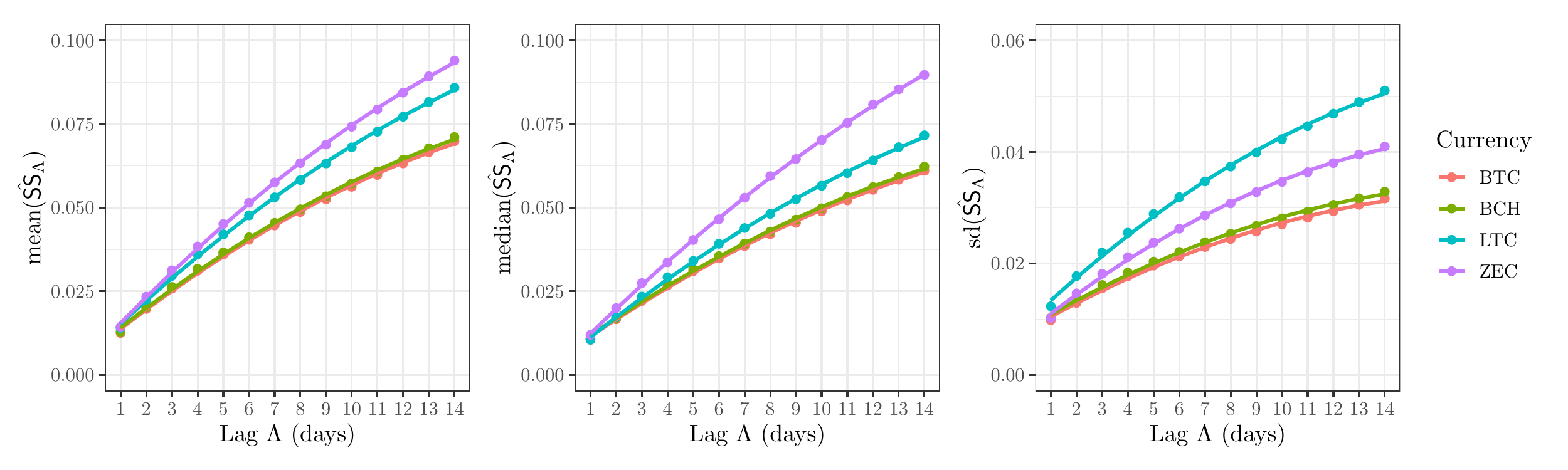}
  \caption{Fitted trends for mean, median, and standard deviation of stake shift.}
  \label{fig:stake_shift_trend}
\end{figure*}

Having observed that stake shifts for different lags evolve synchronously
and vary in amplitudes, we next fitted regression models to the computed mean,
median, and standard deviations (Figure~\ref{fig:stake_shift_trend}).
We can observe that estimated values show a clear, strictly monotonic
increasing trend with growing lag. More specifically, we found that quadratic
polynomials capture well the relation between the location/scale estimators and
lag~$\Lag$ (coefficient of determination $R^2 \geq 0.99$).

\subsection{Attributing Selected Stake Shift Spikes}\label{sec:results_spikes}

In order to shed some more light on the real-world actors behind observable
stake shift spikes, we selected the top five $\Lag = 1$ spikes in each currency
and attributed them to real-world identities using the Chainalysis API.
Due to the limited availability of attribution tags, we focus only on the
period between August 1, 2017 and July 31, 2019. Before continuing, we note
that a fully fledged systematic analysis of real-world entities and their
motivation for transferring large amounts is out of scope in this paper.

\begin{figure}
  \centering
  \includegraphics[width=\columnwidth]{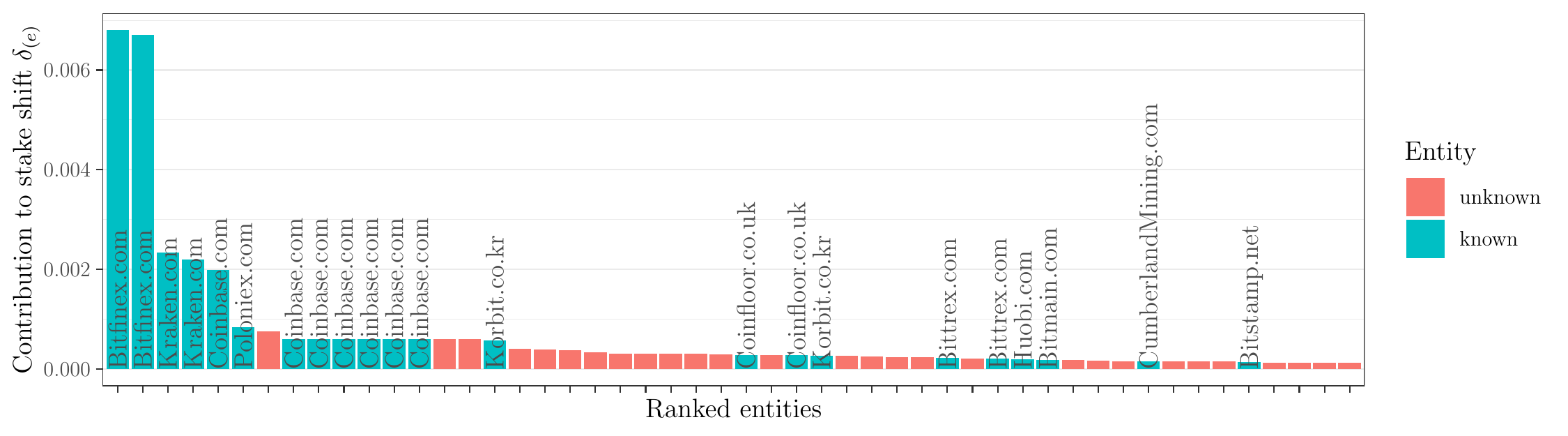}
    \caption{Attributed BTC stake shift spike triggered by Bitcoin Cash fork (2017-08-01)}
  \label{fig:spike_btc_20170801}
\end{figure}

Figure~\ref{fig:spike_btc_20170801} shows the distribution of stake shift
contributions at the spike that occurred during the Bitcoin Cash hard fork
(cf.\ Section~\ref{sec:results_shifts}). We can clearly see that
known exchanges such as Bitfinex, Kraken, Coinbase, and Korbit were the major
entities behind those stake shifts. The largest stake shift was caused by
a transfer from a Bitfinex operated address to some multisig wallet, which is not
a public deposit address but known to be operated by Bitfinex as well. This suggest
that this spike represents a major hot-to-cold wallet transfer. However, it remains
unclear why this co-occurs with the Bitcoin Cash hard-fork date.

We also attributed the top five Bitcoin Cash and Litecoin spikes and see that
exchanges play a major role in stake shifts, however to a lesser extent than
in Bitcoin. In the selected Litecoin spike the identity of involved entities is
unknown. However, we note that only limited attribution tags are available for
that currency. For further details on intra-spike stake shift distributions,
we refer to the plots in the Appendix of this paper
(Figures~\ref{fig:spikes_attributed_btc}--\ref{fig:spikes_attributed_ltc}).

The underlying cause and motivation for being involved in a major stake shift
is not always apparent. Possible reasons are migration of funds between hot
and/or cold wallets, or institutional investors taking a serious long position.
Summarizing the results, we can conclude that, at least in established
cryptocurrencies such as Bitcoin, a small number of real-world entities --
usually exchanges -- may account for major stake shifts in cryptocurrency
ecosystems.

\section{Discussion}\label{sec:discussion}

\heading{Key Findings.}

Our analysis of stabilized stake shift presented in
Sections~\ref{sec:results_shifts} and~\ref{sec:shifts_model}
leads us to the following conclusions:

\begin{itemize}

\item 
  The two main observable reasons for extreme stake-shift spikes are hacks and
  migration of funds to different wallets.
  Large stake shifts resulting from hacks are clearly problematic for a
  proof-of-stake based cryptocurrency, as the entity getting control of
  these funds can be reasonably considered adversarial, with
  unpredictable future actions. 

\item
When considering the levels of adversarial stake ratio that a proof-of-stake
protocol can provably tolerate, one needs to be aware that this threshold is
affected by the assumed maximal stake shift $\sigma$ as per Equation~%
(\ref{eq:alpha}). Our measurements, summarized in Table~%
\ref{tab:stake_shift_stat}, show that depending on the protocol's stake
distribution lag, this effect may decrease the guaranteed resilience bound
by several percent even for lag intervals where the stake shift achieves average
values (as the most extreme example, consider the average stabilized stake
shift for a (hypothetical) two-week lag interval in ZEC, which amounts to
$9.4\%$). Note that, as captured in Figure~\ref{fig:stake_shift_lags_btc} and the standard deviation
values in Figure~\ref{tab:stake_shift_stat}, the stake shift value can deviate
considerably from this average. 
This is particularly noteworthy for protocols that only aim
for the threshold $T=1/3$ in Equation~(\ref{eq:alpha}) such as~\cite{algorand}.

\item
Unsurprisingly, our data confirms that with increasing stake distribution
lag also the corresponding stake shift increases, the precise (empirical)
sublinear dependence is captured in Figure~\ref{fig:stake_shift_trend}.
This advocates for the need to make the stake distribution lag as small as
possible in any future PoS protocol design. More importantly, knowing the
exact slope of this function (and hence the price being paid for longer
stake distribution lag in terms of increased expected stake shift) allows
the designers of existing and future proof-of-stake protocols to weigh
these costs against the benefits of longer lag intervals, leading to more
informed design decisions.

\item
Our results empirically support the natural assumption that high stake shift
    mostly appears at the beginning of the lifetime of a cryptocurrency, and
    hence older, more established cryptocurrencies experience lower average and
    median stake shift for a given lag interval, as well as less occurrences of
    extreme stake shift spikes. This observation allows for some optimism on the
    side of PoS-protocol designers, as the role of stake-shift-related weakening of the
    proven security guarantees should diminish during the lifetime of the
    system. On the other hand, the initial vulnerability of a new, bootstrapping
    PoS cryptocurrency could be prevented for example by the ``merged staking''
    mechanism discussed in~\cite{sidechains}.
\end{itemize}

Additionally, our investigation of the extremal stake-shift spikes conducted in
Section~\ref{sec:results_spikes} results in the following observations:

\begin{itemize}
\item The spikes motivated by migration of funds can be assumed to be often
      triggered by a single entity, we conjecture that the main reason of
      these transfers was moving the considerable funds to a more secure,
      multisig-protected wallet. In such cases, it is natural to assume that
      the funds are controlled by the same party after the transfer, making
      these spikes benign from the perspective of our considered PoS scenario. 
\end{itemize}

\heading{Limitations.}

The main limitation of our results with respect to the question motivating our
investigation lies in the imperfections of clustering techniques and incompleteness of
attribution tags linking entities to real-world identities (despite
using the best currently known). Having a
better understanding of which keys are controlled by the same real-world entity
would give us a more precise picture of the experienced stake shift. However, it
appears likely that more realistic clustering would lead to more keys being
clustered, and hence lower stake-shift estimates. One can thus see our
results as reasonable upper bounds of these quantities.

\heading{Future Work.}

One clear area of future work is to devise new and better address-clustering
and attribution data sharing techniques.
On top of that, it might be interesting to expand our investigation in time and
considered cryptoassets. After more data is available, future studies should
also include assets or currencies built on top of PoS protocols. Such studies
should also investigate the role of exchanges, which typically hold major
stakes and might become important players in a PoS-based consensus. This is
particularly interesting for PoS protocols where coins must be explicitly
``staked'' to participate in the consensus, and hence the total participating
stake is typically much smaller than the overall stake. 
Finally, it would be interesting to perform a more careful and
detailed investigation of the activity behind the five considered major stake shift
spikes, as well as other unusually large spikes uncovered by our work.

\section*{Acknowledgments}
We thank Patrick McCorry for reviewing and commenting on the final draft, and
our AIT colleagues Hannes Koller and Melitta Dragaschnig for insightful
discussions regarding the Apache Spark implementation.
Work on this topic is supported inter alia by the European Union's Horizon 2020
research and innovation programme under grant agreement No.\ 740558 (TITANIUM)
and the Austrian FFG's KIRAS programme under project VIRTCRIME (No.\ 860672).

\bibliographystyle{splncs}
\bibliography{crypto,abbrev3,pubs}

\appendix

\section{Additional Figures}
\label{sec:appendix}

\begin{figure*}
  \centering
  \includegraphics[width=\textwidth]{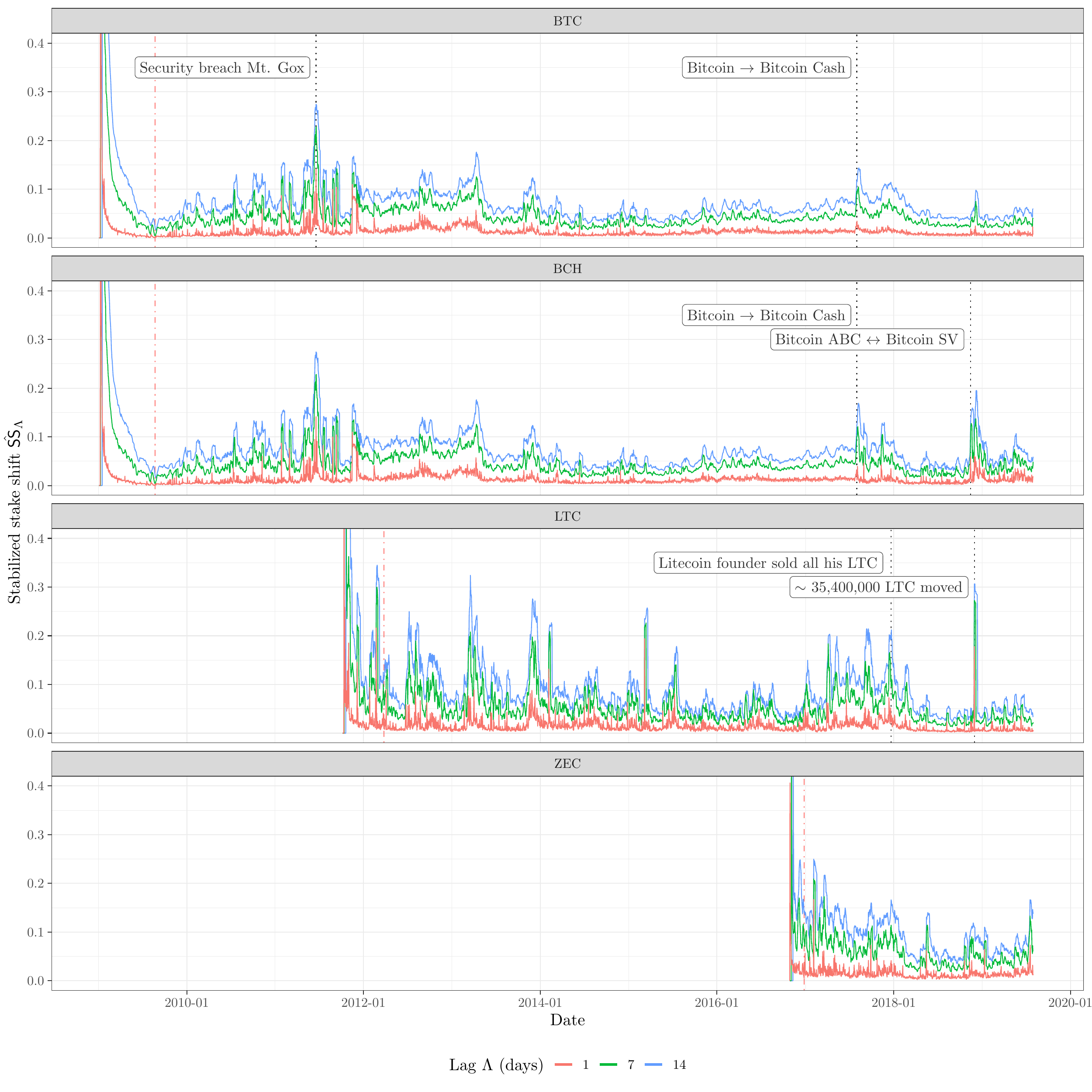}
  \caption{Stake shift for all analyzed cryptocurrencies
           (stake distribution lag $\Lag$: 1, 7, and 14 days).}
  \label{fig:stake_shift_lags}
\end{figure*}

\begin{figure*}
  \centering
    {\includegraphics[width=\textwidth]{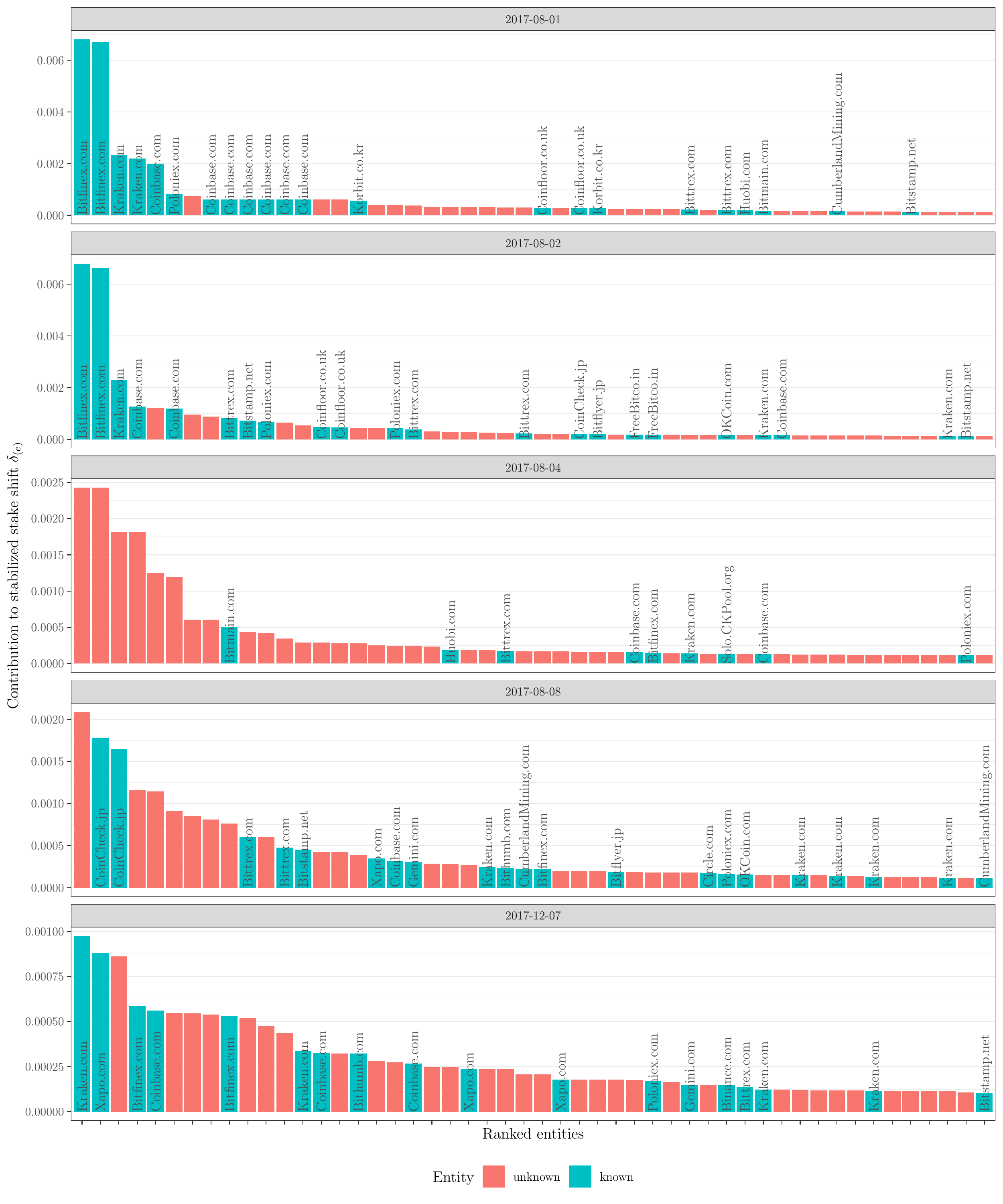}}%
  \caption{Attributed spikes for BTC}
  \label{fig:spikes_attributed_btc}
\end{figure*}

\begin{figure*}
  \centering
    {\includegraphics[width=\textwidth]{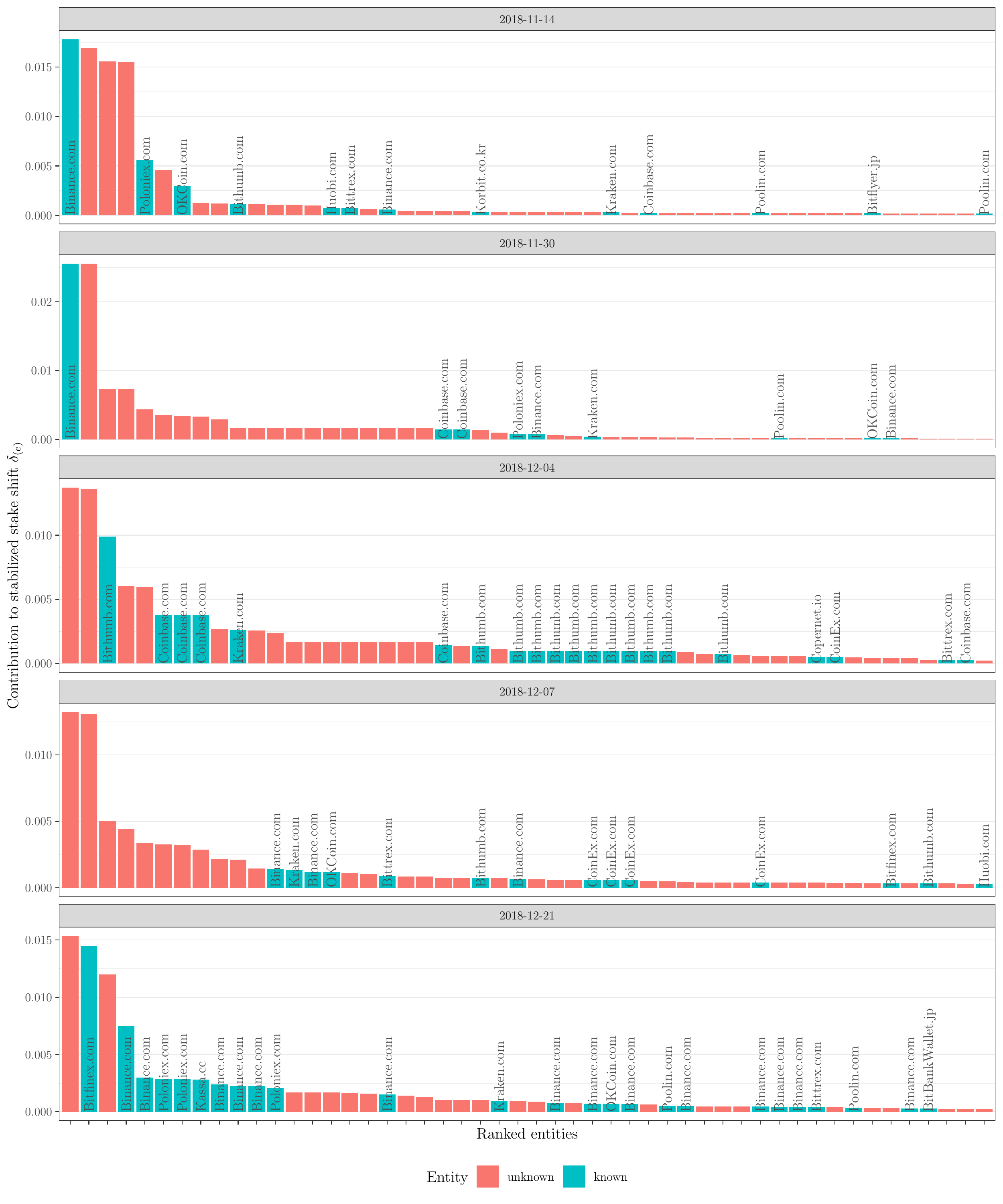}}%
  \caption{Attributed spikes for BCH}
  \label{fig:spikes_attributed_bch}
\end{figure*}

\begin{figure*}
  \centering
    {\includegraphics[width=\textwidth]{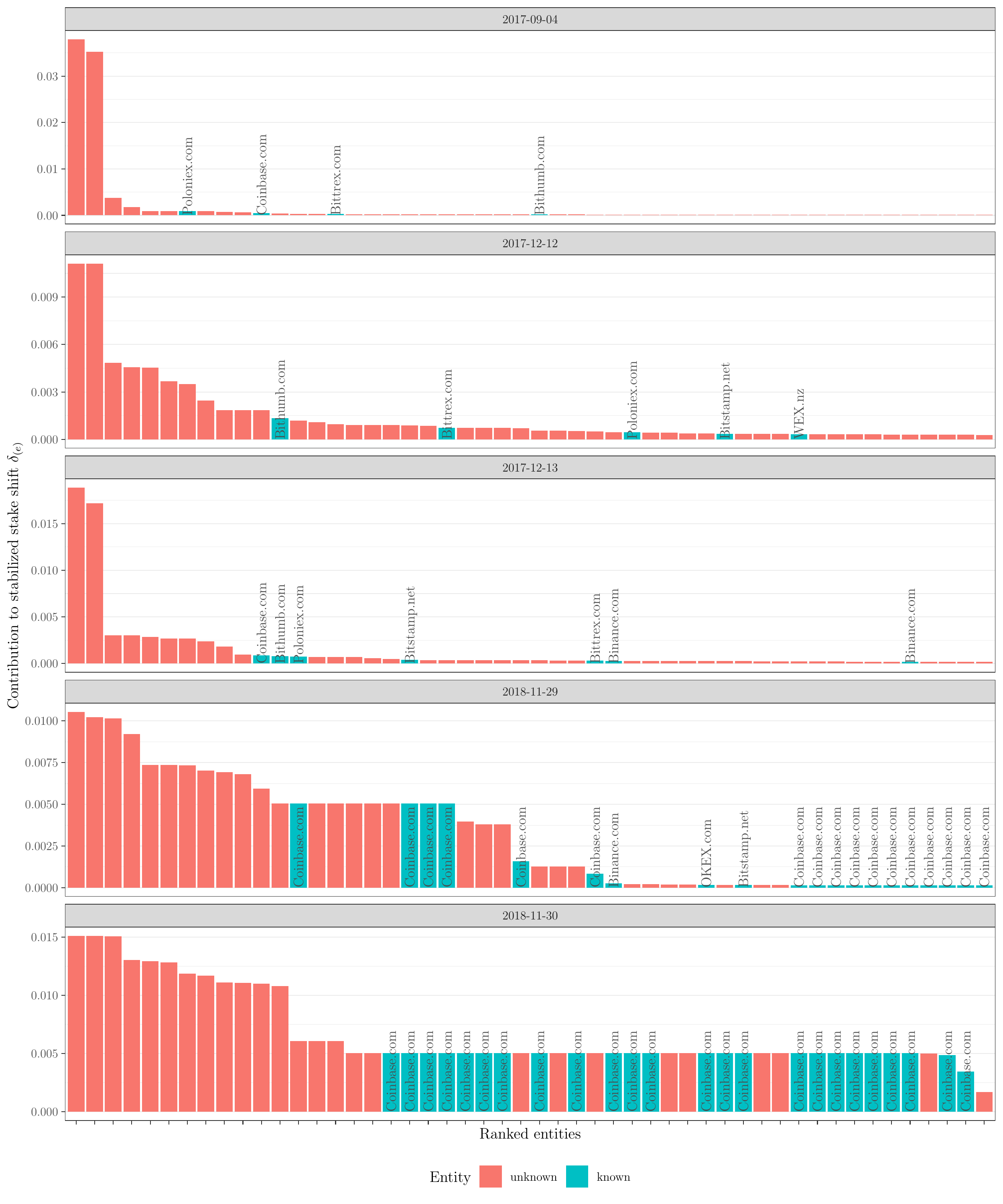}}%
  \caption{Attributed spikes for LTC}
  \label{fig:spikes_attributed_ltc}
\end{figure*}

In this section, we provide additional plots for visual inspection
of our findings reported in Section~\ref{sec:results}.
Figure~\ref{fig:stake_shift_lags} depicts the evolution of stake
shifts for Bitcoin (BTC), Bitcoin (BCH), Litecoin (LTC), and Zcash (ZEC).
Afterwards, in Figures~\ref{fig:spikes_attributed_btc}--\ref{fig:spikes_attributed_ltc}
we present more details on the contributions of real-world
actors to the top-five spikes in each currency within the past two years.


\end{document}